\documentclass[twocolumn,showpacs,graphicx,amsmath,amssymb]{revtex4}
\usepackage{graphicx}

\begin{document}

\title{Properties making a chaotic system 
a good Pseudo Random Number Generator}
\author{Massimo Falcioni, Luigi Palatella, Simone Pigolotti} 
\affiliation{Dipartimento di Fisica and Center for Statistical 
Mechanics and Complexity - INFM,  Universit\`a di Roma "La Sapienza", 
P.le A.Moro 2, Rome 00185, Italy}
\author{Angelo Vulpiani}
\affiliation{Dipartimento di Fisica and Center for Statistical 
Mechanics and Complexity - INFM, Universit\`a di Roma "La Sapienza", 
P.le A.Moro 2, Rome 00185, Italy\\}
\affiliation{INFN, Sezione di Roma ``La Sapienza''}

\date{\today}
\begin{abstract}
We discuss the properties making a deterministic algorithm suitable to
generate a pseudo random sequence of numbers: high value of
Kolmogorov-Sinai entropy, high-dimensionality of the parent dynamical
system, and very large period of the generated sequence.  We propose
the multi dimensional Anosov symplectic (cat) map as a Pseudo Random
Number Generator. We show what chaotic features of this map are useful
for generating Pseudo Random Numbers and investigate numerically which
of them survive in the discrete state version of the map. Testing and
comparisons with other generators are performed.
\end{abstract}
\pacs{05.45.Jn, 05.45.Pq, 05.40.-a}

\maketitle

\section{Introduction}

In most scientific uses of numerical computations, e.g. Montecarlo
simulations and molecular dynamics, it is necessary to have a series
of independent, identically distributed (i.i.d.) continuous random
variables $x(1), x(2),\dots, x(n)$ with assigned single variable
probability density function (pdf) $P(x(i))$. Of course, it is enough
to have i.i.d. random variables $\{x(t)\}$ uniformly distributed in
the interval $[0,1]$, since a suitable change of variable $y=g(x)$ may
generate numbers $\{y(i)\}$ with any pdf $\tilde{P}(y)$.

Let us call perfect random number generator (RNG) a process producing
i.i.d. variables uniformly distributed in $[0,1]$. One can produce
perfect RNG only using nondeterministic physical phenomena, e.g. the
decay of radioactive nuclei or the arrival on a detector of cosmic
rays.

A more practical way is to use a computer that produces a
``random-looking'' sequence of numbers, by means of a recursive rule.
Let us call Pseudo Random Number Generator (PRNG) an algorithm
designed to mimic a random sequence on a computer.  This issue is far
from being trivial; in Von Neumann words: {\em ``Anyone who
considers arithmetical methods of producing random digits is, of
course, in a state of sin'' }\cite{vonneumann}.  The two unavoidable
problems are the following:
\begin{itemize}
\item[$a$)] Numerical algorithms are deterministic.

\item[$b$)] They deal with discrete numbers.

\end{itemize}
The limitations arising from these properties can be analyzed using the
language and the tools of dynamical systems theory. In the following, we
anticipate how these remarks translate in this framework and the main issues
of the entropic characterization of PRNG's (these issues are discussed in
detail in Sec. \ref{due}):

\begin{itemize}

\item[$a_1$)] since the algorithm is deterministic, the Kolmogorov -
  Sinai (KS) entropy ($h_{KS}$) is finite. The sequence $\{x(i)\}$ can
  not be ``really random'', i.e. with an infinite KS entropy, because
  the deterministic dynamical rule constrains the outputs that are
  near in time and supplies us with a maximum of $\log_2 (e^{h_{KS}})$
  random bits per unit time. This limitation would be present also in
  a hypothetical computer able to work with real numbers.

\item[$b_1$)] Since any deterministic system with a finite number of
  states is periodic, any sequence produced by an algorithm working
  with discrete numbers must be periodic, possibly after a transient:
  therefore, not only $h_{KS} < \infty$, but $h_{KS}=0$. The
  computer-implemented system can be only pseudo-chaotic.
 
\end{itemize}

Firstly, we consider point $(a)$.  After the seminal work of Lorenz
and H\'enon \cite{lorenz,henon} (to mention just two of the founders
of the modern theory of chaos), it is well established that also
deterministic systems may have a time evolution that appears rather
``irregular'' with the typical features of genuine random
processes. This evidence opened a debate on the possibility to
distinguish between noisy and chaotic deterministic dynamics.
Following the work of Takens \cite{takens}, the so-called embedding
techniques have been developed to extract qualitative and quantitative
information from a time series. An initial enthusiasm was due to that
the use of the embedding method (via delayed coordinates) allows, at
least in principle, the determination of quantities like dimensions,
$h_{KS}$ and Lyapunov exponents. People believed that, after
determining the KS entropy of a data sequence, one would know the true
nature (deterministic or stochastic) of the law generating the series.
It is now rather clear that there are several limitations in the use
of this technique \cite{ENNE1}: for instance, the number of points
necessary for the phase-space reconstruction increases exponentially
with the dimension of the system \cite{ENNE2}. Thus, due to the
finiteness of the datasets, it is not possible to perform an entropic
analysis with an arbitrary fine resolution, i.e. to compute the
$\epsilon$-entropy $h(\epsilon)$ for very small values of
$\epsilon$. This fact severely restricts the possibility to
distinguish between signals generated by different rules, such as
regular (high dimensional) systems, deterministic chaotic systems, and
genuine stochastic processes. Although the above result may appear
negative, it allows a pragmatic classification of the stochastic or
chaotic feature of the signal, according to the dependence of the
$\epsilon$-entropy on $\epsilon$ and this yields some freedom in
modeling systems \cite{ENNE3}. As a relevant example of a
representation of a deterministic system in term of stochastic
processes, we mention the fully developed turbulence \cite{ENNE4}.
Turbulent systems are high dimensional deterministic chaotic systems
and therefore $h(\epsilon) \simeq h_{KS}$ for $\epsilon \lesssim
\epsilon_{c}$, where $\epsilon_c \rightarrow 0$ as the Reynolds number
$Re \rightarrow \infty$, while $h(\epsilon) \sim \epsilon^{-\alpha}$
for $\epsilon \gtrsim \epsilon_c$. The fact that in certain stochastic
processes $h(\epsilon) \sim \epsilon^{-\alpha}$ can be useful for
modeling purposes: for example, in the so-called synthetic turbulence,
one introduces suitable multi-affine stochastic processes with the
correct scaling properties of the fully developed turbulence.

In this paper we want to discuss about the opposite strategy, i.e. to mimic
noise with deterministic chaotic systems.  Let us summarize the starting
points of our approach to use a deterministic chaotic system as a PRNG:
\begin{enumerate}
\item
Since in any deterministic system $h(\epsilon) \simeq h_{KS}$ for
$\epsilon \lesssim \epsilon_{c}$ with $(\ln \epsilon_{c}) \sim -h_{KS}$,
one should work with a very large $h_{KS}$ \cite{ENNE5}. In this way the true
(deterministic) nature of the PRNG becomes apparent only at a very
high resolution.

\item
The outputs $\{x(t)\}$ of a perfect RNG, when observed at resolution
$\epsilon$, supply $\log_2 (1/\epsilon) \propto h(\epsilon)$ random
bits per iteration. In order to observe the behavior $h(\epsilon) \sim
\ln(1/\epsilon)$ for $\epsilon \gtrsim \epsilon_c$ in a deterministic
algorithm, it is necessary that the time correlation is very weak. We
will discuss how this property may be achieved by taking as output a single
variable of a high-dimensional chaotic system.

 \end{enumerate}

It is not difficult to satisfy point (1), while request (2) is less
obvious. Anosov systems \cite{anosov} are natural candidates to fulfill it,
having invariant stationary measure and very strong chaotic properties.

A third point has to be added, dealing with the problem $(b)$ and its
consequence $(b_1)$. 

Up to here we were considering the chaotic properties of systems with
continuous phase space. Quantities like the $h_{KS}$ and the
$\epsilon$-entropy have an asymptotic nature, i.e. they are related to
large time behavior. However, there are situations where the system
is, strictly speaking, non chaotic ($h_{KS}=0$) but its features
appear irregular to a certain extent.  Such property (denoted with the
term pseudo-chaos \cite{pigo, chirikov, boffetta}) is basically due to
the presence of long transient effects \cite{ENNE6}.

As noted above, the use of a computer discretizes the phase space of a
dynamical system, canceling (at least) its asymptotic chaotic
properties.  However, if the period of the realized sequence is long
enough, the effects related to points $(1)$ and $(2)$ reasonably survive as a
chaotic transient. According to this observation, we add a third
request:
\begin{itemize}
\item[3.]
the period of the series generated by the computer (i.e. with a
state-discretization of the deterministic system) must be very large.
\end{itemize}

Point (3) is really tough: as far as we know, for a generic
deterministic system with $\mathcal{M}$ discrete states, there are no
general methods to determine {\em a priori} the length of the periodic
orbits.  A nice result, based on probabilistic considerations,
suggests that the period $\mathcal{T} \simeq \mathcal{M}^{1/2}$
\cite{coste}, although strong fluctuations are present. The use of
high-dimensional systems may be a natural solution also for this
problem: calling $M$ the number of states along each of the $d$
dimensions, the typical period $\mathcal{T} \simeq M^{d/2}$ grows very
fast by increasing $M$ and $d$.

Whatever the mechanism for producing the pseudo-chaotic transient, the
mere fact that the sequence is periodic implies that it is possible to
obtain equidistributed words only up to a length $\bar{m}=O(\ln
\mathcal{T})$. Thus, long time correlations among outputs of a
generator can not be detected by the standard entropic analysis. We
will show that a high dimensional chaotic system provides outputs
which are not correlated even looking at time delays greater than
$\bar{m}$. In particular we will discuss the connection between
correlation functions and the spectral test for random sequence
\cite{coveyou, marsaglia_pnas, knuth} showing that the outputs of the
high-dimensional cat map have zero $n$-points correlation functions,
when $n$ is less or equal to the dimensions of the map.

In Section \ref{due}, we describe the entropic properties of PRNGs
currently used, underlying both the mechanisms involved in PRNGs. In
Section \ref{duebis} the algorithms used to test the generators are
described.  In Section \ref{tre} we study the properties of the
multi-dimensional Arnol'd's cat map and in Section \ref{quattro} we
propose its discrete version as a PRNG. Section \ref{conclu} is
devoted to conclusions and perspectives.

%
%

\section{Entropy and good PRNG}\label{due}

First of all, we briefly recall some basic notion on the
$\epsilon$-entropy \cite{boffetta}. Consider the variable
$\mathbf{x}(t)\in\mathbb{R}^d$ representing the state of a
$d$-dimensional system, and introduce the new variable
\begin{equation}\label{embed}
\mathbf{y}^{(m)}(t)=(\mathbf{x}(t),\mathbf{x}(t+1),\dots,\mathbf{x}(t+m-1))
\in\mathbb{R}^{md}.
\end{equation}
Of course, $\mathbf{y}^{(m)}$ corresponds to a trajectory in a time
interval $m$. Then, the phase space is partitioned in cells of linear
size $\epsilon$ in each of the $d$ directions. Since the region where
a bounded trajectory evolves contains a finite numbers of cells, each
$\mathbf{y}^{(m)}(t)$ defined in (\ref{embed}) can be coded into a word of
length $m$ out of a finite alphabet:
\begin{equation}
\mathbf{y}^{(m)}(t) \rightarrow W_\epsilon^{(m)}(t)=
(i(\epsilon,t),i(\epsilon,t+1),\dots,i(\epsilon,t+m-1))
\end{equation}
where $i(\epsilon,t+j)$ labels the $\epsilon$-cell containing
$\mathbf{x}(t+j)$. Assuming that the sequence is stationary and ergodic, from
the time evolution of $\mathbf{y}^{(m)}(t)$ the probabilities
$P(\{W_\epsilon^{(m)}\})$ are computed, and one defines the block entropies of size $\epsilon$:
\begin{equation}
H_m(\epsilon)=-\sum_{\{W_\epsilon^{(m)}\}} P\left(W_\epsilon^{(m)}\right)
\ln\left(P(W_\epsilon^{(m)})\right).
\end{equation}
Finally one introduces the $\epsilon$-entropy
$h(\epsilon)$:
\begin{equation}
h(\epsilon)=\lim_{m\rightarrow \infty} h_m(\epsilon)
\end{equation}
where $h_m(\epsilon)=H_{m+1}(\epsilon)-H_m(\epsilon)$ represents the
$\epsilon$-block entropy growth at word length $m$.  In a rigorous
approach, all partitions into elements of size smaller than $\epsilon$
should be taken into account, and then $h(\epsilon)$ is defined as the
infimum over all these partitions \cite{infimum}.  The KS entropy can
be identified as the limit $\epsilon\rightarrow 0$:
\begin{equation}
h_{KS}=\lim_{\epsilon\rightarrow0} h(\epsilon).
\end{equation}

In a deterministic chaotic system, one has $h_{KS}<\infty$, in a
regular motion $h_{KS}=0$, while for a random process with continuous
states $h_{KS}=\infty$. For some stochastic processes, it is possible
to give an explicit expression for $h(\epsilon)$ \cite{kolmogorov}.
For instance, for a stationary Gaussian process with spectrum
$S(\omega)\sim\omega^{-(1+2\alpha)}$ with $0<\alpha<1$ one has
\begin{equation}
h(\epsilon)\sim\epsilon^{-1/\alpha}
\end{equation}
while for i.i.d. variables whose pdf is continuous in a bounded domain
(e.g. independently distributed variables in $[0,1]$) one has:
\begin{equation}
h(\epsilon)\sim\ln \left ( \frac{1}{\epsilon} \right )
\end{equation}
Of course, letting aside the problem of the periodicity induced by the
discrete nature of the states, a PRNG is good when its $h_{KS}$ is
very large, such that the uncertainty on the ``next'' outcome is
larger and the deterministic constraints appear on scales smaller than
an $\epsilon_c$ defined by: $(\epsilon_c)^d \sim e^{-h_{KS}}$.

On the other hand, in data analysis, the space where the state vector
$\mathbf{x}$ lives is unknown and typically in experiments only a
scalar variable $u(t)$ is measured. Therefore, in order to reconstruct
the original phase space, one uses the vector
\begin{equation}
\mathbf{y}^{(m)}(t)=(u(t),u(t+1),\dots,u(t+m-1))
\in\mathbb{R}^{m};
\end{equation}
that is another way to coarse grain the phase space. In this case,
{\it i.e.} looking only at one variable, the maximum scale where the
Kolmogorov-Sinai entropy may be revealed is given by $\epsilon_{c1}
\sim e^{-h_{KS}}$, that is much smaller than $\epsilon_{c}$, for large
$d$ \cite{ENNE5}. Moreover, the single-variable words length that is necessary to
consider, in order to detect $h_{KS}$, must be greater than $d$.
This effect is harmful from the perspective of data analysis, but is 
welcome here.    

We also note that in any series of finite length $T$, it
is not possible to have a good statistics of $m$-words at resolution
$\epsilon$ if $T\lesssim e^{mh(\epsilon)}$. Therefore, for almost all
the practical aims, i.e. for finite $\epsilon$ and finite size of the
sequence, a chaotic PRNG with very high $h_{KS}$ has entropic
properties indistinguishable from those of a perfect RNG.

\subsection{High entropy PRNG}

Several PRNGs are indeed discrete state versions of high entropic dynamical
systems. A popular example is the multiplicative congruential method:

\begin{equation}\label{LCGD}
z(t+1)=a z(t) \quad \mod  M
\end{equation}
where $z(t)$, $a$ and $M$ are integers, with $M\gg a \gg 1$.  In the
following the term ``map'' will denote a dynamical system with
continuous state space, and the term ``automaton'' a system with
discrete state space (we always assume a discrete time).  To avoid
confusion, we will use the symbols $z(t), w(t), \mathbf{z}(t),
\mathbf{w}(t)$ only for discrete dynamical variables (also vectorial)
and $x(t), y(t), \mathbf{x}(t),\mathbf{y}(t)$ for real dynamical
variables.  Eq.(\ref{LCGD}) corresponds to the chaotic map:
\begin{equation}\label{congrdyn}
x(t+1)=a x(t) \quad \mod 1
\end{equation}
where $x(t)=z(t)/M$. It is easy to see that the system
(\ref{congrdyn}) has a uniform invariant p.d.f. in $[0,1]$ and
$h_{KS}=\ln a$. Therefore, only looking at $\epsilon\lesssim
\epsilon_c \sim 1/a$, one can capture the deterministic nature of
the PRNG \cite{ENNE5}.  

It is worthwhile to stress that the chaotic features of the automaton
are apparent only if one observes the system (\ref{LCGD}) after a
coarse-graining procedure, namely with $\epsilon \gg 1/M$. Below this
level of observation, the system keeps a loose trace of the chaotic
features of its continuous precursor and, at the maximal resolution,
at the first time step the block entropy already assumes its maximum
value $H_m(1/M)\approx \ln \mathcal{T}$ for all $m \geq 1$
independently on the value of $h_{KS}$. This happens because we are
observing the complete state of the (one-dimensional) automaton and
suggests again that a suitable attitude is to extract partial
information from high-dimensional systems.

In the next subsection we show an alternative way, based on the
high-dimension effect, to produce random number (up to a given word
length) with an automaton, even at the finest resolution achievable.


\subsection{High dimension PRNG}\label{angelo}

It is known that non-chaotic high dimensional systems may display a
long irregular regime as a transient effect \cite{ENNE6}.  In this
subsection, we show that, with a proper use of a transient irregular
behavior, also systems with a moderate $h_{KS}$ may successfully
generate pseudo-random sequences: in these cases, one observes a
transient in the block $\epsilon$-entropies $H_n(\epsilon)$,
characterized by a maximal (or almost maximal) value of the slope
$H_n(\epsilon)/n$, and then a crossover to a regime with the slope of
the true $h_{KS}$ of the system.
 
The most used class of PRNGs using this property are the so-called lagged
Fibonacci generators \cite{green}, which correspond to the following map:
\begin{equation}\label{fibonacci}
x(t)=ax(t-\tau_1)+bx(t-\tau_2) \qquad \mod 1
\end{equation}
where $a$ and $b$ are $O(1)$ and $\tau_1<\tau_2$.

\begin{figure}[!ht]
\includegraphics[width=8.3 cm]{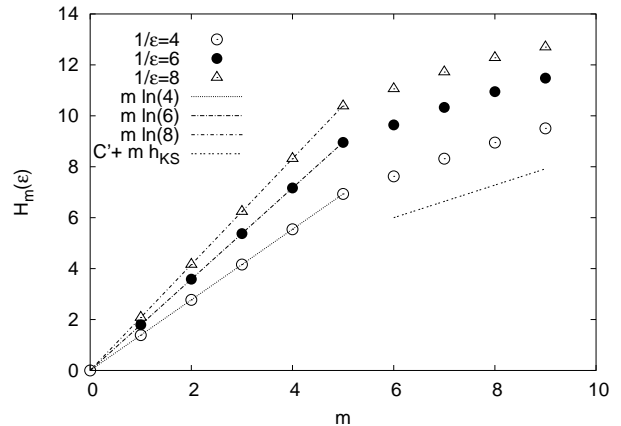}
\caption{\label{fig_fibo} The $\epsilon$ block-entropy for the
Fibonacci map with $\tau_1=2$, $\tau_2=5$ , $a=b=1$ and different values of
$\epsilon$. The change of the slope from $\ln(1/\epsilon)$ to
$h_{KS}$ is clearly visible for $m=\tau_2=5$. }
\end{figure}

Notice that Eq.(\ref{fibonacci}) can be written in the form
\begin{equation}\label{fibo1}
\mathbf{y}(t) = \mathbb{F}\mathbf{y}(t-1)
\end{equation}
where $\mathbb{F}$ is a $\tau_2 \times \tau_2$ matrix of the form
\begin{equation}
\mathbb{F} = 
\left (
\begin{array}{ccccc}
0 & \dots & a & \dots & b \\
1 & 0 & 0 & \dots & 0\\
0 & 1 & 0 & \dots & 0\\
\dots & \dots & \dots & \dots & \dots \\
0 & \dots & \dots & 1 & 0\\ 
\end{array}
\right )
\end{equation}
showing explicitly that the phase space of (\ref{fibonacci}) has
dimension $\tau_2$.  It is easy to prove that this system is chaotic
for each value of $a,b\in \mathbb{N}$, with $a,b>0$. The KS-entropy
does not depend on $\tau_1$ and $\tau_2$ and is of the order $\simeq
\ln(ab)$; this means that to obtain high values of $h_{KS}$ we are
forced to use large values of $a,b$; nevertheless, the lagged
Fibonacci generators are used with $a=b=1$.  For these values of the
parameters $e^{-h_{KS}}\approx 0.618$ and $\epsilon_{c1}$ is not
small. This implies that the determinism of the system should be
detectable also with a large graining. Despite these considerations,
these generators work rather well: the reason is that the $m$-words,
built up by a single variable ($y_1$) of the $\tau_2$-dimensional
system (\ref{fibo1}), have the maximal allowed block-entropy,
$H_m(\epsilon) = m \ln(1/\epsilon)$, for $m < \tau_2$, so that:
\begin{equation}\label{ordinata_origine}
H_m(\epsilon) \simeq \left \{
\begin{array}{ccc}
m \ln \left ( \frac{1}{\epsilon} \right ) & \mathrm{for} & m \lesssim
\tau_2\\ \tau_2 \ln \left ( \frac{1}{\epsilon} \right ) +
h_{KS}(m-\tau_2) & \mathrm{for} & m \gtrsim \tau_2
 \end{array}
\right .
\end{equation}
Eq.(\ref{ordinata_origine}) has the following interpretation: though the
``true'' $h_{KS}$ is small, it can be computed only for very large
value of $m$. Indeed, by observing the one-variable $m$-words, which
corresponds to an embedding procedure, before capturing the dynamical
entropy one has to realize that the system has dimension
$\tau_2$, and this happens only for words longer than
$\tau_2$. Fig. \ref{fig_fibo} shows $H_m(\epsilon)$ for $\tau_1=2$,
$\tau_2=5$ and different values of $\epsilon$.

The importance of the transient behavior of $H_m$ has been underlined
by Grassberger \cite{grassberger} who proposed another quantity beyond
the KS entropy: the ``effective measure of complexity'', namely 
\begin{equation}
C = \sum\limits_{m=1}^{\infty} m (h_{m-1}-h_m).
\end{equation}
From the above definiton, it follows that for large $m$, the block
entropies grow as:
\begin{equation}
H_m \simeq C + m \, h_{KS}.
\end{equation}
For trivial processes, e.g. for Bernoulli schemes or Markov chain of
order $1$, $C=0$ and $h_{KS}>0$, while in a periodic sequence
$h_{KS}=0$ and $C \sim \ln(\mathcal{T})$. In the case of Fibonacci
map, for small $\epsilon$,
\begin{equation}
C = \tau_2 \left [ \ln \left ( \frac{1}{\epsilon} \right ) - h_{KS} \right ]
\approx \tau_2 \ln \left ( \frac{1}{\epsilon} \right ).
\end{equation} 
For large $\tau_2$ (usually values $O(10^2)$ are used) $C$ is so huge
that only an extremely long sequence of the order $\exp(\tau_2)$
(likely outside the capabilities of modern computers) may reveal that
that the ``true'' KS entropy is small.

Let us now discuss the behavior of the discrete Fibonacci generator:
\begin{equation}
z(t) = a z(t - \tau_1) + b z(t - \tau_2) \quad \mod M
\end{equation}
where $z(t) \in [0,M-1]$ and $M \gg \tau_2$. The parameters $\tau_1$,
$\tau_2$ and $M$ are chosen in order to have a period as long as
possible. Number-theoretical arguments \cite{knuth} allow to choose
these parameters such that the period of the orbit is maximum
$\mathcal{T}=M^{\tau_2}-1$.

When the period is maximum,
for $\epsilon \geq 1/M$ one has:
\begin{equation}\label{3regimi}
H_m(\epsilon) \simeq \left \{
\begin{array}{ccc}
m \ln \left ( \frac{1}{\epsilon} \right ) & \mathrm{for} & m \lesssim
\tau_2\\ \tau_2 \ln \left ( \frac{1}{\epsilon} \right ) +
h_{KS}(m-\tau_2) & \mathrm{for} & \tau_2 \lesssim m \lesssim m^* \\
\tau_2 \ln(M) & \mathrm{for} & m \gtrsim m^*
 \end{array}
\right .
\end{equation}
where 
\begin{equation}
m^* =  \frac {\tau_2}{h_{KS}} \left [ \ln \left (
\frac{1}{\epsilon}\right)  -\ln M + h_{KS} \right ].
\end{equation}
When $\epsilon=1/M$ we have $m^*=\tau_2$, the second regime in
Eq.(\ref{3regimi}) disappears and the block entropy behavior is
independent of $h_{KS}$. Still, as for the continuous case, if
$\tau_2$ is large one observes only the pseudo-chaotic transient
\begin{equation}
H_m(\epsilon) \approx m
\ln \left ( \frac{1}{\epsilon} \right ).
\end{equation}

Summarizing, systems with high values of $h_{KS}$ or with high
dimension, produce sequences having entropic properties rather close
to those of a perfect RNG, for two different reasons. In the first
case, the large $h_{KS}$ allows to use a small $\epsilon$ (but large
enough to achieve a proper coarse-graining): in such a way, the
$\epsilon$-entropy coincides with that of a perfect RNG. In the second
case, the high dimensionality of the system prevents the entropic
analysis to reveal the asymptotic value of $h_{KS}$ before the end of
a long transient behavior mimicking a complete random system.
   

In conclusion, a deterministic PRNG has good entropic properties for long (but
finite) sequences if $h_{KS}$ or $C$ are large. In Section \ref{tre} we will
propose a multi-dimensional cat map as a PRNG having both these properties.


\section{Tests for PRNGs}\label{duebis}

Several techniques have been developed in order to test ``how random''
is a given sequence of numbers. These algorithms are available in
easy-to-use software packages collecting dozens of different tests,
like, for example, the {\em DieHard} \cite{diehard} and the {\em NIST}
\cite{nist} batteries. Many of them compute the frequency of some
words $f(z(j),z(j+1),\dots,z(j+n))$ made up of $n$ consecutive outputs
of the generator and compare it to the theoretical probability in the
random case. Examples are: the frequency and block-frequency tests
(computing $f(z(j))$, i.e. $m=1$), Poker test looking for words with
$m=5$ corresponding to the Poker hands (e.g. Full house: 00011, four a
kind: 00001), template tests checking the occurrences of some ($\simeq
10^2$) words with length $m=10-12$.

It is worth stressing that all these benchmarks are automatically passed if
the block $\epsilon$-entropy, is maximal for words of length $m$,
namely
\begin{equation}\label{maxentr}
H_m(\epsilon) = m \ln(M).
\end{equation}
with $\epsilon=1/M$ where $M$ is the number of symbols produced by the
generator. When $H_m$ is maximal, it is impossible to distinguish the
output from a truly random sequence by looking only at $m$ consecutive
symbols.  The reason for introducing many tests, instead of looking
only at the entropy, is that a little departure from a constant word
frequency implies a correction in the entropy that is only quadratic in
the deviation (this is a simple consequence of the fact that
the value of the entropy in Eq. (\ref{maxentr}) is the maximum
achievable).  Thus, it is really difficult to numerically observe
unbalances in the frequency of some specific words by studying only
the block entropies.

\subsection{The spectral test}

The entropic analysis is a very powerful tool from a theoretical point
of view but it presents a major limitation: in a phase space with a
finite number of states the block entropy cannot grow more than $\ln
\mathcal{T}$, where $\mathcal{T}$ is the period of the orbit. This
fact essentially fixes an upper bound on the number of possible
equidistributed words and on their length. Even with the longest
periods available in current used algorithm, the bound on the length
is of order $10^2 - 10^3$. On the other hand, computer simulations
often necessitate of a large amount of random numbers, and long-term
relationships among these numbers can be sources of hard-to-discover
biases \cite{ferrenberg,mertens1,mertens2}.  Therefore, less severe
tests than the entropic one are needed.  One can ask that the
correlations among different outputs (or, more generally, among
different {\em functions} of the outputs), vanish even when the
outputs are at a time distance greater than the scale where
Eq.(\ref{maxentr}) ensures the equidistribution of the words.  On this
timescale, numbers should {\em appear} random, as far as one is
interested on statistical observable, even if, looking at the whole
sequence, later numbers are completely determined by previous
ones. Indeed, according to the entropic analysis, the knowledge of
approximately $\ln \mathcal{T}$ consecutive outputs permits to
determine exactly the discrete starting condition of the system and
consequently to predict the whole sequence, removing any randomness
from it.

The main tool to analyze the property of correlation functions is the
{\em spectral test}. We start defining the frequency
$f(z(t_1),\dots,z(t_n))$ of the word $(z(t_1),\dots,z(t_n))$ as we
made for the Kolmogorov entropy, where now $t_i$'s are generic times
and $z(t_i)$'s are not in general consecutive outputs. The spectral
test is the multi-dimensional Fourier transform of
$f\left(z(t_1),\dots,z(t_n)\right)$
\begin{equation}
\hat f(s_1, s_2, .....,s_n) = \left \langle \exp \left (\frac{2\pi i}{M}
\sum\limits_j s_j z(t_j) \right ) \right \rangle,
\end{equation}
where $s_j \in [0,M-1] $, $M$ is the number of the discrete states
and $\langle \dots \rangle$ denote the average over the trajectory.
For true random numbers:
\begin{equation}\label{randomspectra}
\hat f(s_1, s_2, .....,s_n)=\delta_{s_1,0}\delta_{s_2,0}\dots \delta_{s_n,0}
\end{equation}
for any choice of $n$ and of the time lags $t_i$'s. Values of the
function $\hat f(s_1, s_2, .....,s_n)$ significantly different from
$0$ denote wave vectors of probability density fluctuations in the
lattice $z(t_1),z(t_2)\dots z(t_n)$. These fluctuations can be safely
neglected only when their characteristic length scale (which we assume
to be the inverse of the modulus of the wave vector) is much smaller
than the maximum precision one is interested in. Many generators
(i.e. the linear congruential class of generators) produce numbers
that ``fall mainly in planes'' \cite{coveyou, marsaglia_pnas, knuth},
and the presence of these planes is detected by the spectral test.

The importance of the spectral test is related to the fact that
analytical or semi-analytical methods \cite{knuth,dieter} allow for
a fast calculation for simple systems. Furthermore, since any $L^2$
function can be written as a Fourier series, condition
(\ref{randomspectra}) implies the vanishing of any correlation of up to 
$n$ functions of time-delayed variables:
\begin{eqnarray}\label{anycorrelation}
\langle g_1(z(t_1))g_2(z(t_2))\ldots g_n(z(t_n))\rangle = \nonumber
\\ = \langle g_1(z(t_1)) \rangle \langle g_2(z(t_2))\rangle\dots \langle
g_n(z(t_n)) \rangle
\end{eqnarray}
for every $g_i \in L^2$.

\section{The cat-map as a random number generator}\label{tre}

Recently some authors \cite{shchur} proposed the use of the Arnol'd
cat map as a PRNG. We will briefly recall the properties of this map
and then propose a multi-dimensional version with $N$ coupled maps,
showing that this generalization gives rise to very good statistical
properties.  In particular, it has both the properties analyzed in
Section \ref{due}  giving maximal $\epsilon$-entropy, namely it
possesses a high value of $h_{KS}$ and it is a high-dimensional system.
We will see that this system has also very good properties from
the point of view of correlation functions.

The 2-dimensional Arnol'd's cat map \cite{arnold} is a symplectic
automorphism on a torus satisfying the property of Anosov systems
\cite{ott}, namely it is everywhere hyperbolic and has a positive
Kolmogorov entropy. The map reads
\begin{equation}\label{gatto2D}
\left ( 
\begin{array}{c} 
x' \\
y'
\end{array}
\right ) = \left ( 
\begin{array}{cc} 
1 & a \\
b & 1+ab
\end{array}
\right )
\left ( 
\begin{array}{c} 
x \\
y
\end{array}
\right ) \:\: {\rm mod} \:\: 1 ,
\end{equation}
where $a,b \in \mathbb{N}$. 
The standard example given by Arnol'd is obtained with $a=b=1$. 

The multi-dimensional generalization can be written in the following way:
\begin{equation}\label{MDgatto}
\left ( 
\begin{array}{c} 
\mathbf{x'} \\
\mathbf{y'}
\end{array}
\right ) =  \mathbb{M}
\left ( 
\begin{array}{c} 
\mathbf{x} \\
\mathbf{y}
\end{array}
\right ) \:\: {\rm mod} \:\: 1,
\end{equation}
with
\begin{equation}\label{simplettica}
\mathbb{M} = \left ( 
\begin{array}{cc} 
\mathbb{I} & \mathbb{A} \\
\mathbb{B} & \mathbb{I+BA}
\end{array}
\right )
\end{equation}
where $\mathbb{M}$ is a $2N \times 2N$ matrix, $\mathbf{x,y} \in
\mathbb{R}^N$, $\mathbb{I}$ is the $N \times N$ identity matrix and
$\mathbb{A, B}$ are symmetric $N\times N$ matrices with natural
entries in order to obtain a continuous mapping.  It easy to see that
the evolution law given by Eq.(\ref{simplettica}) is symplectic,
indeed one can write Eq.(\ref{simplettica}) and Eq.(\ref{MDgatto}) as
a canonical transformation
\begin{equation}
\mathbf{x} = \frac{\partial S(\mathbf{x'},\mathbf{y})}{\partial \mathbf{y}}, 
\quad
\mathbf{y'} = \frac{\partial S(\mathbf{x'},\mathbf{y})}{\partial \mathbf{x'}}
\end{equation}
where the generating function is given by
\begin{equation}
S(\mathbf{x'},\mathbf{y}) = \sum\limits_{j=1}^{N} x'_j y_j - \frac{1}{2} 
 \sum\limits_{j,k=1}^{N} ( y_j A_{jk}y_k + x'_j B_{jk} x'_k).
\end{equation}
It can be shown that when ${\rm Tr}(\mathbb{M}) > 2N $ map
(\ref{MDgatto}) is an Anosov system with uniform invariant
measure. The output of our generator will be the first component of
the vector $\mathbf{x}$.  The condition $N > 1$ raises the Kolmogorov
entropy, cancels the correlation and increases the length of the
periodic orbits (in the discretized case). We will describe in detail
these three aspects in the following.

First of all, according to Pesin identity, the Kolmogorov entropy of this
system is equal to the sum of the positive Lyapunov exponents:
\begin{equation}
h_{KS}=\sum_{\lambda_i>0}\lambda_i.
\end{equation}
A $d$-dimensional hyperbolic symplectic system possesses exactly $d/2$
positive Lyapunov exponents; if they are of the same order of
magnitude, the Kolmogorov entropy grows proportional to the number of
dimensions. Notice that the Kolmogorov entropy may also be raised by
simply taking the matrix $\mathbb{A, B}$ with very large entries. This
method, however, produces only an increase in the entropy which is
logarithmic in the size of the entries.  Of course for the system
(\ref{MDgatto} - \ref{simplettica}) the $\lambda_i$ are easily
obtained from the eigenvalues $\alpha_i$ of $\mathbb{M}$: $\lambda_i =
\ln \vert\alpha_i\vert$.

For the 2D cat map we compute an approximate value of the
$\epsilon$-entropy, obtained as $H_4(\epsilon)-H_3(\epsilon)$ varying
$\epsilon$ and the parameters $a,b$. In order to highlight the
practical limitations of PRNG we study the discrete version of
Eq.(\ref{gatto2D}) with $M=2^{20}$ possible values of $x$ and $y$ (see
the next section for details).

\begin{figure}[!ht]
\includegraphics[width=8.3 cm]{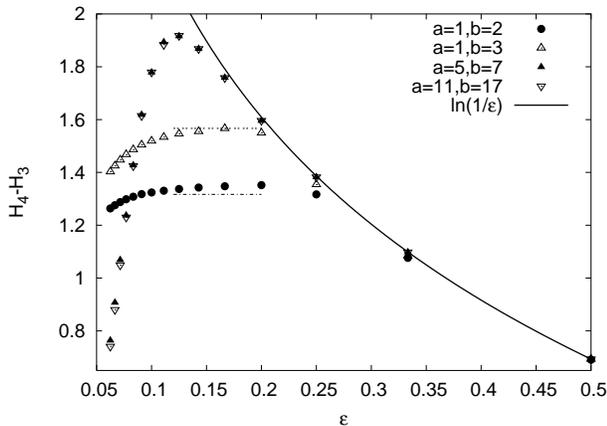}
\caption{\label{fig_entro} $H_4(\epsilon)-H_3(\epsilon)$ for the 2D
cat map of Eq.(\ref{gatto2D}) with $M=2^{20}$, different values of $a,b$ as a
function of $\epsilon$. The horizontal lines indicate the $h_{KS}$ values}
\end{figure}

 As one can observe in Fig.\ref{fig_entro}, both the standard problems
of PRNGs appear. Indeed the figure shows that decreasing the value of
$\epsilon$ we observe a ``plateau'' around the value of $h_{KS}$.  At
lower values of $\epsilon$, there is an abrupt decrease due to the
periodic nature of the map.  Nevertheless it seems that if we use the
map as a generator of a number of symbols $\simeq 1/\epsilon$ with
$\epsilon>\epsilon_c$ we are, with a good approximation, near the
value corresponding to a theoretical RNG, given by
$h(\epsilon)=-\ln(\epsilon)$. 
Let us note that, when $h_{KS}$ is large enough (curves for $a=5$, $b=7$,
$h_{KS} \approx 3.61$ and $a=11$, $b=17$, $h_{KS}\simeq 5.24$), because of the
limited number of allowed states, one does not observe the plateau
$h(\epsilon)\sim h_{KS}$.

Let us study the properties of the time correlation of the outputs.
The following result holds: 
{\em Let $\mathbf{e}_1$ be the
2N-dimensional vector $(1,0,0 \dots)$. If the vectors
$(\mathbb{M}^T)^{t_1} \mathbf{e}_1, (\mathbb{M}^T)^{t_2} \mathbf{e}_1,
\dots,(\mathbb{M}^T)^{t_{2N}}\mathbf{e}_1$ are linearly independent,
then one has:}
\begin{equation}\label{spectralresult}
 \left \langle \exp \left (2\pi i
\sum\limits_{j=1}^{2N} s_j x_1(t_j) \right ) \right \rangle =
\delta_{s_1,0}\delta_{s_2,0}\dots \delta_{s_{2N},0}. 
\end{equation}
{\em Furthermore, the independence of the vectors is ensured for any choice of
  the time delays $t_i$'s if the matrix $\mathbb{M}$ has real positive and
  non-degenerate eigenvalues and the vector $\mathbf{e}_1$ has non-zero
  component on all the eigenvectors.} For the proof see Appendix A.

The practical meaning of this result is the following: we observe only
the variable $x_1$, keeping the remaining $2N-1$ variables hidden, and
we study its correlation functions. In this way correlation functions
involving up to $2N$ different times vanish,
i.e. Eq.(\ref{anycorrelation}) holds for $n \leq 2N$, because the
contributions due to different values of the hidden variables cancel
out in the averaging.

The result of Eq.(\ref{spectralresult}) can be taken as one of the strongest
characterization of a finite random sequence: as we said, word
equidistribution can hold only up to a value of $\bar{n} \simeq\ln T$ where
$T$ is the length of the sequence.  Indeed, some authors \cite{compagner}
define as random sequence of length $ T$ one containing all the possible words
up to length $\bar{n}$. On the other hand, Eq.(\ref{spectralresult}) is a
generalization of that condition: for consecutive time delays $t_j=j$ the two
properties are equivalent, while for generic values of the $t_j$'s it ensures
long-range independence of the outputs, without asking an exponential number
of equiprobable words in the sequence.

The validity of the property of (\ref{spectralresult}) in the
discrete case will be subject of careful analysis in the following
section. Here, we just point out that, even in the continuous case,
this property is not shared by some of the dynamical systems used for
generating random numbers. For example, let us recall the Fibonacci map 
\begin{equation}
x(t)=a x(t-\tau_1)+b x(t-\tau_2) \quad \mod 1
\end{equation}
with $a,b \in \mathbb{N}$ with $a,b>0$ and $\tau_2>\tau_1$. 
It is straightforward to show that the correlation function
\begin{equation}
\left \langle \exp \{ 2 \pi i (s_1 x(t) + s_2 x(t-\tau_1) 
+ s_3 x(t-\tau_2)) \} \right \rangle
\end{equation}
is not equal to zero for the vector $\mathbf{s}=(s_1,s_2,s_3)=(1,-a,-b)$.
Thus, even if the dimension of the phase
space $\tau_2$ may be very high, it is sufficient the
three-points correlation function to unveil the deterministic nature
of the system.

Therefore, when the dimension of the cat
map, $2N$, is equal to the dimension of the Fibonacci generator
$\tau_2$, both the systems guarantee that words of length $2N$ are
equidistributed. However, the main advantage of the multi-dimensional cat map
is that also the words made up of $2N$ non-consecutive symbols are
equidistributed.  This property does not hold for Fibonacci generators
and in some case this can lead to serious problems.  A famous example
is the ``Ferrenberg affair'' \cite{ferrenberg}: persistence in
binary Fibonacci generators gives misleading results in Montecarlo
simulation. This problem is well analyzed in the framework of
information theory in \cite{mertens1}.


In the next section we numerically study the discrete version of the
multidimensional cat map and we check whether the good statistical
properties of the system survive in this case.

\section{Numerical analysis and test of the multidimensional cat automaton}
\label{quattro}

A digital computer cannot handle real numbers.  What a computer
really calculates is a finite-digit dynamics that can be represented
as a dynamics on integers. We will consider in the following the
multidimensional cat automaton, namely
\begin{equation}\label{discreta}
\left ( 
\begin{array}{c} 
\mathbf{z'} \\
\mathbf{w'}
\end{array}
\right ) =  
\left ( 
\begin{array}{cc} 
\mathbb{I} & \mathbb{A} \\
\mathbb{B} & \mathbb{I+BA}
\end{array}
\right )
\left ( 
\begin{array}{c} 
\mathbf{z} \\
\mathbf{w}
\end{array}
\right ) \:\: {\rm mod} \:\: M,
\end{equation}
where, as usual, $\mathbb{A,B}$ have natural entries, $z_i, w_i \in
[0,1,\dots,M-1]$.

The first problem in passing from the continuous to the discrete case
is that the system has a finite number of state $M^{2N}$ and
consequently it must be periodic and it is no more truly chaotic.  The
optimal condition is that there is only one orbit covering all the
states but the origin ${\bf z=w=0}$ (which is a fixed point) thus
obtaining a period $\mathcal{T}=M^{2N}-1$.  A peculiar feature of cat
maps is that periodic orbits of the continuum system have rational
coordinates \cite{vivaldi}.  Consequently, the orbits of the
discretized version of the map are completely equivalent to periodic
orbits of the continuous system with coordinates $z_i/M, w_i/M$. As a
corollary, being the map invertible, periodic orbits do not have any
transient: every state is recurrent.

\begin{figure}[!h]
\includegraphics[width=8.3 cm]{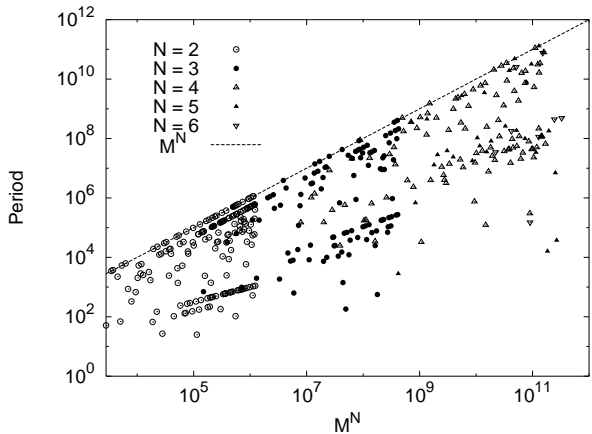}
\caption{\label{fig_periodo} The period of the orbits for the multidimensional
cat map, with different values of $N$, as a function of $M^N$ (only the prime
values of $M$ are considered). The straight lines refer to the probabilistic
argument $\mathcal{T} \simeq M^N$.}
\end{figure}

Unfortunately the cat map has typically many orbits, and the great
majority of them are of the same length. A theoretical analysis of
these orbits have been made \cite{vivaldi} in the 2-dimensional
case. For generic dynamical system, arguments based on random maps
suggest that the average length of the orbits $\mathcal{T}$ should be
roughly the square root of the total number of states \cite{henon}: in
our case $\mathcal{T}\approx M^N$. This scaling have been numerically
observed in typical chaotic dynamical systems \cite{grebogi}. In
Fig. \ref{fig_periodo}, we show the length, $\mathcal{T}$, of the
orbits as a function of $M$ and $N$. Despite the large fluctuations,
one retrieve the expected qualitative scaling and, more importantly,
the periods seem to be independent on the initial condition; this
suggests that several symmetry operations exist mapping one orbit into
another one of the same length. Since the choice of $M$ is critical in
determining the length of the orbits, we restrict ourselves to prime
number values of $M$, in order to avoid the presence of trivial
invariant sub-lattices generated by the divisors of $M$. Notice also
that data are plotted as a function of $M^N$: the lengths show the
correct exponential growth with $N$, at least in a statistical sense.
We stress that in Fig. \ref{fig_periodo} we show the result for $M$
not very large ($<10^3$), from the observed behavior one can say that
for $M \sim 10^9$ the period should be extremely large.

Unfortunately we have no theoretical control over the period, and wild
fluctuations are present when $M$ varies; therefore it is better to
choose a value of $M, N, \mathbb{A,\: B}$ and directly check the value
of $\mathcal{T}$ or a lower bound. In the following we will consider
the choice $N=3$, $M = 1,001,400,791$ and
\begin{equation}\label{scelta}
\mathbb{A}=
\left ( 
\begin{array}{ccc} 
1 & 1 & 1\\
1 & 3 & 1 \\
1 & 1 & 5
\end{array}
\right )\quad  \quad
\mathbb{B} =  
\left ( 
\begin{array}{ccc} 
7 & 1 & 1\\
1 & 3 & 1 \\
1 & 1 & 9
\end{array}
\right ).
\end{equation}
With these parameters, the hypothesis leading to Eq.(\ref{spectralresult})
hold, furthermore we numerically obtained $\mathcal{T} > 7 \cdot 10^{12}$,
that is a satisfying lower bound for typical simulations.

The very encouraging result we obtained for the correlation functions
in the continuous case is the main reason for the use of the
multidimensional cat automaton as a PRNG, but we have to check whether the
property proven in the previous section holds also in the discrete case.
The choice of (\ref{scelta}) satisfies the hypothesis of the theorem,
namely the eigenvalues are all strictly positive, non-degenerate and
the vector $(1,0,0,0,0,0)$ has non-zero component along all
eigenvectors.

From a general point of view there are two main technical {\em
caveats} when passing from continuous to discrete systems concerning
the theoretical spectral test. In our case, since the orbit do not
cover all the space, it is {\em a priori} impossible to average on the
uniform distribution as discussed in the Appendix for the continuous
state case. Even if we suppose that the orbit is sufficiently
homogeneous that Eq.(\ref{ergodic}) (see the Appendix) is approximately
valid, condition (\ref{linsys}) becomes in the discrete case
\begin{equation}\label{linsysdiscreta}
\sum\limits_{j=1}^{2N} s_j m_{1k}^{(t_j)}=0\:\: {\rm mod}\:\: M 
\quad \forall k=1,\dots, 2N
\end{equation}
 since it is sufficient to consider vectors $(s_1,s_2,\dots,s_{2N})$
in the first Brillouin zone.

These two reasons prevent us to perform the spectral test
using the nice theoretical arguments used for other kind of generators
\cite{knuth,ecuyer} and force us to use numerical simulation.
This constitutes a hard computational task and we perform the test
for low values of $M$ and studying products of the form
\begin{equation}
\hat f(s_1,s_2)=
\left \langle 
\exp \left (\frac{2 \pi i}{M} s_1 z_1(t_1) \right )
\exp \left (\frac{2 \pi i}{M} s_2 z_1(t_2) \right )
 \right \rangle.
\end{equation}
We use $M=1031$ and $N=3$ obtaining a period $\mathcal{T}=274,243,921$ and
letting $s_1,s_2 \in [0,M-1]$. Using an FFT numerical algorithm we check up to
time delays $t_2-t_1\le250$ that for all values, but $s_1=s_2=0$, $|\hat
f(s_1,s_2)|<10^{-5}$.  With a lower number of states, $M=127$,
$\mathcal{T}=1,016,188$, we compute also the three-point spectral test,
obtaining always values compatible with the inverse of the square root of the
period $\mathcal{T}$. This suggests that the periodic orbits look like a
finite statistical sample of the continuum equilibrium distribution, as far as
one studies only few-point correlation functions. An important remark is that
a solution $(s^{*}_1,s^*_2,\dots,s^*_{2N})$ of the Diophantine
Eq.(\ref{linsysdiscreta}) implies that $\hat f(s^{*}_1,s^*_2,\dots,s^*_{2N}) =
1$ independently on the stationary distribution and, consequently, on the
period $\mathcal{T}$. This means that the low values observed in the numerical
spectral test exclude the possibility of a solution in
Eq.(\ref{linsysdiscreta}).

In order to look for any other possible bias, we also apply the {\em
NIST} battery to test our multidimensional cat automaton, with the
parameters of Eq.(\ref{scelta}), for generating $10^3$ binary string
of $0$'s and $1$'s of length $10^6$; all tests performed with the
recommended parameters have been passed.

\section{Conclusions}\label{conclu}

In this paper we show how, using properties of high dimensional
deterministic chaotic systems, it is possible to generate a good
approximation of a random sequence, in spite of unavoidable constraints
of deterministic algorithms running on digital computers.

Summarising, we have two possible mechanisms to obtain good PRNGs
using deterministic systems: very high KS entropy, and ``transient
chaos'' with a large finite-time $\epsilon$ entropy, due to the high
dimensionality of the system.  We propose the multi-dimensional cat
map as a PRNG having both these properties.  Another important example
of system with both the properties is the one proposed by Knuth
\cite{knuth-page}: one iterates the Fibonacci generator
(\ref{fibonacci}) with $M=2^{31}-1$, $\tau_1=37$ and $\tau_2=100$,
with this choice the period is extremely large, then the output
sequence is obtained taking the variable in Eq.(\ref{fibonacci}) every
$T$ steps ($T=1009$ or $2009$).  In such a way, for the words of size
up to $\tau_2$ (i.e. extremely huge), the $\epsilon$ entropy is
practically $\simeq \ln (1/\epsilon)$, i.e. like a perfect
RNG. Moreover, even if the simple Fibonacci generator fails the
three-points spectral test, it is harder to find non-null vector in
the spectral test of Knuth's generator, because of the fact that $T$,
$\tau_1$ and $\tau_2$ are relatively prime numbers.  Nevertheless, it
does not seem that a general result like that of
Eq.(\ref{spectralresult}) may be easily extended to this PRNG.


We suggest that the multi-dimensional cat map is suitable for
generating random number sequence. The main advantage if compared with
other generators, is the factorization of all $n$-times, with $n<2N$,
correlation function due to the high dimensionality of the system and
the presence of hidden variables. This result is rigorously true (also
in the case $n=2N$) in the continuous system; numerical checks show
that this property survives in the discrete case.  Moreover, this map
has a large value of the KS entropy giving good entropic properties at
non-zero, but small, $\epsilon$.

A disadvantage of this method is that we can not predict analytically
the period given the parameters or, equivalently, write a condition on
the parameters in order to obtain the maximum period.  However,
probabilistic arguments \cite{coste}, confirmed by numerical check,
show that the period increases exponentially with $N$, therefore with
a proper choice of the parameters we achieve extremely large
periods. An analytical criterium to predict the length of the period
could pave the way to the application of multi dimensional cat maps as
high quality PRNGs.

\begin{acknowledgments}
We thank M. Cencini and F. Cecconi for useful remarks and a careful reading of
the manuscript, and the support by MIUR - COFIN03 ``Complex Systems and
Many-Body Problems'' (2003020230).
\end{acknowledgments}

\section*{APPENDIX A}

Consider the system of Eqs. (\ref{MDgatto}) and (\ref{simplettica}).
In this Appendix we give the proof of the following proposition:

{\em Let $\mathbf{e}_1$ be the
2N-dimensional vector $(1,0,0 \dots)$. If the vectors
$(\mathbb{M}^T)^{t_1} \mathbf{e}_1, (\mathbb{M}^T)^{t_2} \mathbf{e}_1,
\dots,(\mathbb{M}^T)^{t_{2N}}\mathbf{e}_1$ are linearly independent,
then one has:}
\begin{equation}
 \left \langle \exp \left (2\pi i
\sum\limits_{j=1}^{2N} s_j x_1(t_j) \right ) \right \rangle =
\delta_{s_1,0}\delta_{s_2,0}\dots \delta_{S_{2N},0}. 
\end{equation}
{\em Furthermore, the independence of the vectors is ensured for any choice of
  the time delays $t_i$ if the matrix $\mathbb{M}$ has real positive and
  non-degenerate eigenvalues and the vector $\mathbf{e}_1$ has non-zero
  component on all the eigenvectors.}

Proof. Since the system under study is ergodic and its invariant 
measure is uniform, we can write the average in Eq.(\ref{spectralresult}) as:
\begin{equation}\label{ergodic}
\int dx_1\dots dx_{2N} \exp \left (2\pi i
\sum\limits_{j=1}^{2N} s_j x_1(t_j) \right ).
\end{equation}
Here, with an abuse of notation, we call $x_j$ the components of both
the $\mathbf{x}$ and the $\mathbf{y}$ vectors, i.e. $x_{j+N} \equiv y_j$, $j
= 1,\dots,N$.  Let us also call $m_{jk}^{(t)}$ the elements of the
matrix $\mathbb{M}^t$. We can rewrite the previous expression in the
following way:
\begin{equation}\label{intaverage}
\int dx_1\dots dx_{2N} \exp \left (2\pi i
\sum\limits_{j,k=1}^{2N} s_j m_{1k}^{(t_j)} x_k\right )
\end{equation}
Notice that we do not take care about the modulus since the $s_j$ are integers
and the complex exponential is periodic.
When integrating over the $x_i$'s, the result is zero for every value of the
$s_i$'s, excluding the values that are solution of the linear system:
\begin{equation}\label{linsys}
\sum\limits_{j=1}^{2N} s_j m_{1k}^{(t_j)}=0 \qquad \forall k=1\dots 2N
\end{equation}
that yield $1$ as a result of Eq.(\ref{intaverage}). Since $s_j=0\
\forall j$ is a trivial solution for the linear system, it is
sufficient to show that this solution is unique to demonstrate
Eq. (\ref{spectralresult}). In particular, by Cramer's rule, it is
sufficient to show that $\det \mathbb{G}\neq 0$, where $\mathbb{G}$ is
the matrix of coefficients of the linear system (\ref{linsys}), namely
\begin{equation}
g_{kj}= m_{1k}^{(t_j)}.
\end{equation}
Notice that the column of the matrix $\mathbb{G}$ are
constituted by the component of the vectors $(\mathbb{M}^T)^{t_j}\mathbf{e}_1$;
this means that the condition of having $\det \mathbb{G}\neq0$ 
is equivalent to require that the vectors $(\mathbb{M}^T)^{t_j}\mathbf{e}_1$
are linearly independent. This completes the first part of the proof.

Now, we show how Eq. (\ref{linsys}) has always a unique solution when
the matrix $\mathbb{M}$ has positive and non-degenerate eigenvalues
$\lambda_k$, and the vector $\mathbf{e}_1$ has non-zero component
along all the eigenvectors, In this case, we rewrite the matrix
$\mathbb{G}$ in the eigenvector basis, obtaining
\begin{equation}
\det \mathbb{G} =  c_1 c_2 \dots c_{2N}
\det \left ( 
\begin{array}{cccc} 
\lambda_1^{t_1}  &  \lambda_1^{t_2} & \dots & \lambda_1^{t_{2N}} \\
\lambda_2^{t_1}  &  \lambda_2^{t_2} & \dots & \lambda_2^{t_{2N}} \\
\dots & \dots & \dots & \dots \\
\lambda_{2N}^{t_1}  &  \lambda_{2N}^{t_2} & \dots & \lambda_{2N}^{t_{2N}} \\
\end{array}
\right )
\end{equation}
where $c_k \neq 0$ for hypothesis is the component of $\mathbf{e}_1$ along the
k-th eigenvector. Notice that the $c_k$'s are real: since the eigenvalues are
real by hypothesis, also the eigenvectors have real components in the natural
basis of $\mathbb{R}^{2N}$. The proof is a reduction {\em ad absurdum}: let us
suppose that $\det \mathbb{G}=0$, this implies that there exist a linear
combination of the columns satisfying
\begin{equation}\label{polin1}
\sum_j b_j \lambda_k^{t_j}=0 \quad \forall k=1\dots 2N.
\end{equation}
Calling $P(z)$ the polynomial
\begin{equation}
P(z)\equiv \sum_j b_j x^j
\end{equation} 
Eq.(\ref{polin1}) implies that the polynomial $P(z)$ has $2N$ distinct
real positive roots being, by hypothesis, all eigenvalues positive and
non-degenerate. Then, by Descartes sign rule, it must have at least
$2N$ sign changes in the coefficients, but this is impossible, since
$P(z)$ has just $2N$ terms different from $0$. Thus, $\det \mathbb{G}$
is necessarily different from zero for any possible choice of the time
delays; this completes the proof.

\end{document}